\documentclass[english,showpacs,showkeys]{revtex4-2}
\usepackage[utf8]{inputenc}
\usepackage{color}
\usepackage{babel}
\usepackage{array}
\usepackage{units}
\usepackage{amsmath}
\usepackage{graphicx}
\usepackage[bookmarks=true,bookmarksnumbered=false,bookmarksopen=true,bookmarksopenlevel=1,
 breaklinks=false,pdfborder={0 0 0},pdfborderstyle={},backref=false,colorlinks=true]
 {hyperref}
\hypersetup{
 pdfauthor={S. G. Salnikov},
 citecolor=blue,urlcolor=blue,linkcolor=blue}

\makeatletter

\providecommand{\tabularnewline}{\\}

\usepackage{bm}

\makeatother

\begin{document}
\title{Meson production in $J/\psi$ decays and $J/\psi\to N\bar{N}\gamma$
process}
\author{S. G. Salnikov}
\email{S.G.Salnikov@inp.nsk.su}

\affiliation{Budker Institute of Nuclear Physics, 630090, Novosibirsk, Russia}
\affiliation{Novosibirsk State University, 630090, Novosibirsk, Russia}
\author{A. I. Milstein}
\email{A.I.Milstein@inp.nsk.su}

\affiliation{Budker Institute of Nuclear Physics, 630090, Novosibirsk, Russia}
\affiliation{Novosibirsk State University, 630090, Novosibirsk, Russia}
\date{\today}
\begin{abstract}
It is shown that an account for the final-state interaction of real
or virtual nucleon and antinucleon produced in the processes $J/\psi\to p\bar{p}\gamma$,
$\psi(2S)\to p\bar{p}\gamma$, $J/\psi\to p\bar{p}\omega$, and $J/\psi\to3\left(\pi^{+}\pi^{-}\right)\gamma$
near the threshold of $N\bar{N}$ pair production allows one to obtain
self-consistent description of these processes. Predictions of our
model are in good agreement with experimental data available. The
proposed potential model also reproduces the corresponding partial
cross sections of $p\bar{p}$ scattering.
\end{abstract}

\maketitle
\noindent
\global\long\def\im#1{\qopname\relax{no}{Im}#1}%

\section{Introduction}

Invariant mass $M$ of a nucleon-antinucleon pair $N\bar{N}$ in the
decay $J/\psi\to N\bar{N}+A$, where $A=\gamma,\rho,\omega,\pi^{0},\eta$,
in the rest frame of $J/\psi$ is determined by the energy $E_{A}$
of particle~$A$, $M^{2}=m_{J/\psi}^{2}-2m_{J/\psi}E_{A}$. Therefore,
measurement of $E_{A}$ allows one to fix the value of~$M$. Anomalous
behavior of the decay probabilities has been observed in the processes
$J/\psi\to p\bar{p}+A$~\citep{Bai2001,Ablikim2009,Bai2003,Ablikim2008,Alexander2010,Ablikim2012,Ablikim2013b}
and $J/\psi\to\textrm{mesons}+A$~\citep{Ablikim2016,Ablikim2023e},
where the invariant mass of produced mesons is close to the double
proton mass, $M\approx2m_{p}$. Usually, this anomalous behavior is
explained by the existence of a family of resonances $X(1835)$, $X(1840)$,
$X(1870)$, and others. However all available experimental data can
be well explained within the approach based on an account for the
interaction of real or virtual nucleon and antinucleon produced in
$J/\psi$ decays (see~\citep{Kang2015,Dmitriev2016a,Milstein2017,Dedonder2018,Dai2018,Yang2023}
and references therein). The same approach explains successfully experimental
data for the cross sections of processes $e^{+}e^{-}\to\text{mesons}$
near the threshold of real $N\bar{N}$ pair production (see~\citep{Haidenbauer2015,Dmitriev2016,Milstein2022c}).

In the approach based on the account for the final-state interaction,
a quark-antiquark pair is produced at small distances $r\sim1/2m_{p}$
and then transforms into a nucleon-antinucleon pair at large distances
$r\sim1/\Lambda_{QCD}$ as a result of hadronization. For small relative
velocity of $N$ and $\bar{N}$, the $N\bar{N}$ interaction may significantly
increase the modulus of wave function $|\psi(0)|$ (here $\psi(0)$
is the value of wave function of $N\bar{N}$ pair at distances $r\sim1/\Lambda_{QCD}$).
Firstly, this happens when there is a loosely bound $N\bar{N}$ state
with the binding energy $\left|\varepsilon\right|\ll\left|\overline{U}\right|$,
$\varepsilon<0$, where $\overline{U}$ is the characteristic value
of the potential $U(r)$ of $N\bar{N}$ interaction. Secondly, there
is no loosely bound state, but a slight increase of the potential
depth results in its appearance. We refer to the latter case as a
virtual state with an energy $\varepsilon\ll\left|\overline{U}\right|$,
$\varepsilon>0$. In both cases, an energy $\varepsilon$ is expressed
in terms of the $N\bar{N}$ scattering length $a$, $\left|\varepsilon\right|=1/m_{p}a^{2}$,
where $\left|a\right|$ is much larger than the characteristic size
$R$ of the potential. Moreover, $a>0$ in the case of a loosely bound
state and $a<0$ for a virtual state.

We refer to the production of a real $N\bar{N}$ pair as an elastic
process. A produced virtual $N\bar{N}$ pair can annihilate into a
system of mesons, we refer to such process as inelastic. The sum of
probabilities of elastic and inelastic processes is the total probability.
The inelastic processes are possible above the threshold of real $N\bar{N}$
pair production as well as below this threshold (due to annihilation
of a virtual pair). Therefore, the anomalous behavior of the probabilities
of decays $J/\psi\to N\bar{N}+A\to\textrm{mesons}+A$ are determined
by the energy dependence of the probability of virtual $N\bar{N}$
production.

The quantum numbers of a nucleon-antinucleon pair are determined by
that of particle $A$ and also by the quantum numbers of mesonic system.
For instance, the main contribution to the probabilities of $J/\psi\to N\bar{N}\gamma(\rho,\omega)$
decays near the $N\bar{N}$ production threshold is given by the production
of $N\bar{N}$ pair with the angular momentum $l=0$, total spin $S=0$,
and charge parity $C=+1$. Then, the isospin of $N\bar{N}$ pair is
$I=1$ in the decay with $\rho$ meson production, isospin $I=0$
with $\omega$ meson production, and the isospin $I$ is not fixed
in the decay $J/\psi\to N\bar{N}\gamma$. In the process $J/\psi\to(6\pi)\gamma$
the $G$-parity of $6\pi$ state is $G=+1$, and the $C$-parity of
$N\bar{N}$ pair in the process $J/\psi\to N\bar{N}\gamma$ is $C=+1$.
Therefore, the contribution to the probability of the decay $J/\psi\to(6\pi)\gamma$
is given by virtual $N\bar{N}$ pair with the isospin $I=0$. Let's
now consider the process $J/\psi\to(6\pi)\pi^{0}$, where $6\pi$
are produced through an intermediate $N\bar{N}$ state. In this case,
the $C$-parity of $N\bar{N}$ pair is $C=-1$, total spin $S=1$,
and the isospin of the pair is $I=1$, as well as the isospin of produced
$6\pi$ system. Therefore, the effective nucleon-antinucleon potentials
in the processes $J/\psi\to N\bar{N}\gamma$ and $J/\psi\to N\bar{N}\pi^{0}$
are different, and the probabilities of the corresponding processes
are also different. This is the reason why a large number of resonances
$X$ has been introduced for interpretation of anomalous behavior
of probabilities in various processes with the meson production.

In the present paper, using the approach based on the account for
the final-state interaction, we perform a self-consistent description
of the processes $J/\psi\to p\bar{p}\gamma$, $\psi(2S)\to p\bar{p}\gamma$,
$J/\psi\to p\bar{p}\omega$, and $J/\psi\to3\left(\pi^{+}\pi^{-}\right)\gamma$.

\section{Results}

In Ref.~\citep{Milstein2017} we successfully described the experimental
data for the energy dependence of $J/\psi\to p\bar{p}\gamma(\omega)$
decays probabilities. Therefore, we can predict the behavior of $J/\psi\to(6\pi)\gamma$
decay probability near the threshold of $N\bar{N}$ pair production.
The probability of the process $J/\psi\to p\bar{p}\gamma$ can be
written as (see Ref.~\citep{Milstein2017})
\begin{align}
 & \frac{d\Gamma_{p\bar{p}\gamma}}{dM}=\frac{p\,k^{3}}{2^{4}\thinspace3\pi^{3}m_{J/\psi}^{4}}\left|\mathcal{G}_{\gamma0}\,\psi^{(0)}(0)+\mathcal{G}_{\gamma1}\,\psi^{(1)}(0)\right|^{2},\nonumber \\
 & k=\frac{m_{J/\psi}^{2}-M^{2}}{2m_{J/\psi}}\,,\qquad p=\sqrt{m_{p}E}\,,\qquad E=M-2m_{p}\,.\label{eq:gamma}
\end{align}
Here $k$ is the photon momentum in the $J/\psi$ rest frame, $p$
is the nucleon momentum in the $N\bar{N}$ center of mass frame, $\mathcal{G}_{\gamma0}$
and $\mathcal{G}_{\gamma1}$ are some energy-independent constants
related to the amplitudes of $N\bar{N}\gamma$ state production at
small distances. The functions $\psi^{(I)}(r)$ are the regular solutions
of the radial Schrödinger equations for $N\bar{N}$ pair with the
corresponding isospin~$I$. The probability of $J/\psi\to p\bar{p}\omega$
decay reads (see Ref.~\citep{Milstein2017})
\begin{equation}
\frac{d\Gamma_{p\bar{p}\omega}}{dM}=\frac{p\,p_{\omega}^{3}}{2^{4}\thinspace3\pi^{3}m_{J/\psi}^{4}}\left|\mathcal{G_{\omega}}\psi^{(0)}(0)\right|^{2},\qquad p_{\omega}=\sqrt{\varepsilon_{\omega}^{2}-m_{\omega}^{2}}\,,\qquad\varepsilon_{\omega}=\frac{m_{J/\psi}^{2}+m_{\omega}^{2}-M^{2}}{2m_{J/\psi}}\,,\label{eq:omega}
\end{equation}
where $\mathcal{G_{\omega}}$ is some constant.

The total probability $\Gamma_{\mathrm{tot}}^{(0)}$ of $J/\psi\to N\bar{N}+\gamma$
and $J/\psi\to N\bar{N}+\gamma\to\textrm{mesons}+\gamma$ processes,
in which real or virtual $N\bar{N}$ pair has the isospin $I=0$,
is expressed via the Green's function $\mathcal{D}^{(0)}(r,r'|E)$
of the radial Schrödinger equation for $N\bar{N}$ pair with quantum
numbers $l=0$, $S=0$, and $I=0$ (see Ref.~\citep{Milstein2017})
\begin{equation}
\frac{d\Gamma_{\mathrm{tot}}^{(0)}}{dM}=-\frac{\left|\mathcal{G}_{\gamma0}\right|^{2}k^{3}}{2^{4}\thinspace3\pi^{3}m_{p}m_{J/\psi}^{4}}\im{{\cal D}^{(0)}\left(0,\,0|E\right)}.
\end{equation}
The probability of elastic process $J/\psi\to N\bar{N}\gamma$, where
$N\bar{N}$ has $I=0$, reads
\begin{equation}
\frac{d\Gamma_{\textrm{el}}^{(0)}}{dM}=\frac{\left|\mathcal{G}_{\gamma0}\right|^{2}p\,k^{3}}{2^{4}\thinspace3\pi^{3}m_{J/\psi}^{4}}\left|\psi^{(0)}(0)\right|^{2}.
\end{equation}
The probability $d\Gamma_{\text{inel}}^{(0)}/dM$ of inelastic decays
$J/\psi\to\textrm{mesons}+\gamma$, in which the system of mesons
has $I=0$, is
\begin{equation}
\frac{d\Gamma_{\text{inel}}^{(0)}}{dM}=\frac{d\Gamma_{\text{tot}}^{(0)}}{dM}-\frac{d\Gamma_{\text{el}}^{(0)}}{dM}\,.\label{eq:InelMassSpectrum}
\end{equation}
Note that the effects of isotopic invariance violation (the proton
and neutron mass difference and the Coulomb interaction of proton
and antiproton) only slightly affect $d\Gamma_{\text{inel}}^{(0)}/dM$.

In a recent work~\citep{Ablikim2023e}, the distribution $d\Gamma_{6\pi}/dM$
over the invariant mass of $3\left(\pi^{+}\pi^{-}\right)$ in the
decay $J/\psi\to3\left(\pi^{+}\pi^{-}\right)\gamma$ has been measured
with high accuracy. To describe the probability of this process, it
is necessary to take into account both the contribution of virtual
$N\bar{N}$ pair with $I=0$ in an intermediate state and the contributions
of mechanisms not related to the annihilation of a virtual $N\bar{N}$
pair. The energy dependence of latter contributions is a smooth function
of $M$ near the threshold of real $N\bar{N}$ pair production, while
the former contribution depends strongly on $M$ in the near-threshold
region. A smooth dependence on $M$ of contributions, which are not
related to the virtual $N\bar{N}$ pair production, can be approximated
using few parameters~\citep{Ablikim2023e}.

Eq.~(\ref{eq:InelMassSpectrum}) predicts the sum of probabilities
of all inelastic processes, and the production of $3\left(\pi^{+}\pi^{-}\right)$
system is only one of possible channels. However, it is natural to
assume that the annihilation amplitude of $N\bar{N}$ pair into mesons
weakly depends on $M$ near the threshold, and it can be considered
a constant. Therefore, the contribution of virtual $N\bar{N}$ annihilation
to the probability of $J/\psi\to3\left(\pi^{+}\pi^{-}\right)\gamma$
decay is proportional to $d\Gamma_{\text{inel}}^{(0)}/dM$. This assumption
was fully justified when describing the anomalous behavior of meson
production cross sections in $e^{+}e^{-}$ annihilation near the threshold
of real $N\bar{N}$ production~\citep{Dmitriev2016,Milstein2017,Milstein2022c}.

Using the experimental data~\citep{Ablikim2023e} we found that the
contribution of background processes in the near-threshold region
can be approximated with good accuracy by a linear function of energy~$E$.
As a result, we describe the distribution $d\Gamma_{6\pi}/dM$ by
the formula
\begin{equation}
\frac{d\Gamma_{6\pi}}{dM}=a+b\,E+c\,\frac{d\Gamma_{\text{inel}}^{(0)}}{dM}\,,\label{eq:6pi}
\end{equation}
where $a$, $b$ and $c$ are some parameters that have been determined
by comparison of our predictions with experimental data.

In order to determine $d\Gamma_{\text{inel}}^{(0)}/dM$ it is necessary
to find the parameters of $N\bar{N}$ interaction potential with quantum
numbers $l=0$, $S=0$, and $I=0$. We have used the experimental
data on the production of $p\bar{p}$ in $J/\psi\to p\bar{p}\gamma$,
$\psi(2S)\to p\bar{p}\gamma$ and $J/\psi\to p\bar{p}\omega$ decays~\citep{Bai2003,Ablikim2008,Alexander2010,Ablikim2012,Ablikim2013b},
and also the results of partial-wave analysis of elastic and inelastic
$p\bar{p}$ scattering data performed by the Nijmegen group~\citep{zhou2012energy}.
In processes $J/\psi\to p\bar{p}\gamma$ and $\psi(2S)\to p\bar{p}\gamma$,
$p\bar{p}$ pairs are produced both with $I=0$ and $I=1$ (see Eq.~(\ref{eq:gamma})).
In the process $J/\psi\to p\bar{p}\omega$, the $p\bar{p}$ pair is
produced with $I=0$ (see Eq.~(\ref{eq:omega})). However, the experimental
data for this decay are limited as compared to the decay $J/\psi\to p\bar{p}\gamma$.
Therefore, we have used the whole set of experimental data listed
above to better fix the parameters of the potential. As a result,
we found not only the effective interaction potential of $N\bar{N}$
with $I=0$, but also with $I=1$.

\renewcommand{\arraystretch}{1.5}
\begin{table}
\begin{centering}
\begin{tabular}{|>{\raggedright}p{2cm}|>{\centering}p{2cm}|>{\centering}p{2cm}|}
\hline 
 & $U^{(0)}$ & $U^{(1)}$\tabularnewline
\hline 
$V\,\textrm{(MeV)}$ & $-92$ & $-24$\tabularnewline
\hline 
$W\,\textrm{(MeV)}$ & $114$ & $89$\tabularnewline
\hline 
$R\,\textrm{(fm)}$ & $1.17$ & $1.06$\tabularnewline
\hline 
\end{tabular}
\par\end{centering}
\caption{Parameters of potentials~(\ref{eq:pot}) of $N\bar{N}$ interaction
in the states with isospin $I=0,1$.}\label{tab:pot}
\end{table}

\begin{figure}
\begin{centering}
\includegraphics[totalheight=5.3cm]{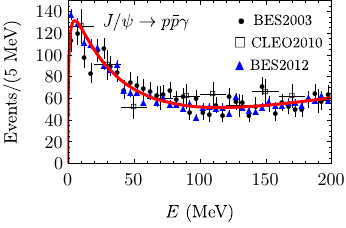}\hfill{}\includegraphics[totalheight=5.3cm]{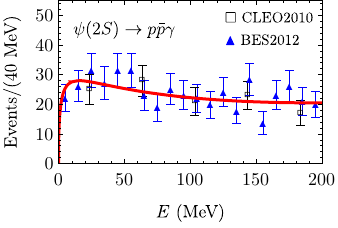}
\par\end{centering}
\begin{centering}
\includegraphics[totalheight=5.3cm]{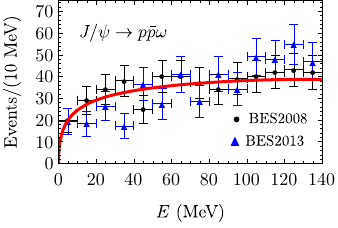}
\par\end{centering}
\caption{Energy dependence of the probabilities of decays $J/\psi\to p\bar{p}\gamma$,
$\psi(2S)\to p\bar{p}\gamma$, and $J/\psi\to p\bar{p}\omega$ in
comparison with experimental data~\citep{Bai2003,Ablikim2008,Alexander2010,Ablikim2012,Ablikim2013b}.
All graphs are normalized to the number of events in the earliest
experiment.}\label{fig:1}
\end{figure}

As shown in Refs.~\citep{Salnikov2023,Salnikov2024}, the behavior
of cross sections in the near-threshold region is determined by a
small number of parameters (scattering lengths, effective ranges of
interaction). Therefore, one can use any convenient parameterization
of the effective potentials, which reproduces the required values
of these parameters. We have used the parameterization of $N\bar{N}$
interaction potentials in states with $l=0$, $S=0$, and $I=0,1$
in the form of rectangular wells
\begin{equation}
U^{(I)}(r)=\left(V^{(I)}-i\,W^{(I)}\right)\cdot\theta(R^{(I)}-r)\,,\label{eq:pot}
\end{equation}
where $V^{(I)}$, $W^{(I)}$ and $R^{(I)}$~--- are some parameters,
and $\theta(x)$ is the Heaviside function. The optical potential
$U^{(I)}(r)$ contains an imaginary part that accounts for annihilation
of $N\bar{N}$ pair into mesons. For such a parameterization one can
obtain the analytical form of the wave functions $\psi^{(I)}(r)$
and the Green's functions $\mathcal{D}^{(I)}(r,r'|E)$. We have (see
Ref.~\citep{Salnikov2024})
\begin{align}
 & \psi^{(I)}(0)=\frac{q\,e^{-ipR^{(I)}}}{q\cos\left(qR^{(I)}\right)-ip\sin\left(qR^{(I)}\right)}\,,\nonumber \\
 & \im{\mathcal{D}^{(I)}(0,0|E)}=\im{\left[q\,\frac{q\sin\left(qR^{(I)}\right)+ip\cos\left(qR^{(I)}\right)}{q\cos\left(qR^{(I)}\right)-ip\sin\left(qR^{(I)}\right)}\right]},\nonumber \\
 & q=\sqrt{m_{p}\left(E-V^{(I)}+i\,W^{(I)}\right)}\,.
\end{align}

The values of potential parameters, that provide the best fit of the
experimental data, are given in the Table~\ref{tab:pot}. Fig.~\ref{fig:1}
shows the comparison of our predictions with experimental data for
$J/\psi\to p\bar{p}\gamma$, $\psi(2S)\to p\bar{p}\gamma$, and $J/\psi\to p\bar{p}\omega$
decays. In all cases, good agreement between the predictions and the
experimental data is evident. We have checked that our model is consistent
with the results of partial-wave analysis of $p\bar{p}$ scattering
data performed by the Nijmegen group~\citep{zhou2012energy}.

The Green's function $\mathcal{D}^{(0)}(r,r'|E)$ for quantum numbers
$l=0$, $S=0$, $I=0$ has poles in the complex energy plane at $E=E_{R}^{(0)}$.
For $W^{(0)}=0$ we obtain $E_{R}^{(0)}=\unit[-2]{MeV}$, which correspond
to subthreshold resonance. For $W^{(0)}=\unit[114]{MeV}$ we have
$E_{R}^{(0)}=\unit[\left(36-57i\right)]{MeV}$, that corresponds to
the unstable bound state, see Ref.~\citep{Badalyan1982}. Relatively
large imaginary part of $E_{R}^{(0)}$ explains absence of a pronounced
peak in the probability of $J/\psi\to p\bar{p}\omega$ decay.

\begin{figure}
\begin{centering}
\includegraphics[totalheight=5.3cm]{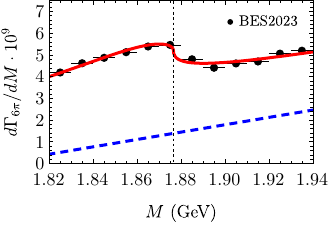}
\par\end{centering}
\caption{The dependence of $J/\psi\to3(\pi^{+}\pi^{-})\gamma$ decay probability
on the invariant mass $M$. The solid line is our predictions for
$d\Gamma_{6\pi}/dM$, the dashed line is the background contribution.
The vertical dotted line indicates the $N\bar{N}$ threshold. Experimental
points are recalculated from Ref.~\citep{Ablikim2023e}.}\label{fig:2}
\end{figure}

Using the obtained potentials and Eq.~(\ref{eq:6pi}), we compare
our predictions with the recent experimental data~\citep{Ablikim2023e}.
To obtain $d\Gamma_{6\pi}/dM$ one should multiply the number of $J/\psi$
decays into $3(\pi^{+}\pi^{-})\gamma$ observed in Ref.~\citep{Ablikim2023e}
by the ratio $\Gamma_{J/\psi}/N_{J/\psi}$, where $\Gamma_{J/\psi}=92.6\,\mbox{keV}$
is the total width of $J/\psi$ meson and $N_{J/\psi}=10087\cdot10^{6}$
is the total number of $J/\psi$ events. Our comparison is shown in
Fig.~\ref{fig:2} for $a=1.4\cdot10^{-9}$, %
\mbox{%
$b=0.017\cdot10^{-9}\,\mbox{MeV}^{-1}$%
}, and %
\mbox{%
$c=7\cdot10^{-4}$%
}. It is seen that our model successfully reproduces the nontrivial
energy dependence of $J/\psi\to3\left(\pi^{+}\pi^{-}\right)\gamma$
decay probability near the $N\bar{N}$ threshold.

\section{Conclusion}

A simple model, based on the account for the final-state interaction,
is proposed for self-consistent description of the processes $J/\psi\to p\bar{p}\gamma$
, $\psi(2S)\to p\bar{p}\gamma$, $J/\psi\to p\bar{p}\omega$ and $J/\psi\to3\left(\pi^{+}\pi^{-}\right)\gamma$
near the threshold of $N\bar{N}$ pair production. It is shown that
the nontrivial energy dependence of the probabilities of these processes
is related to the interaction of real and virtual $N$ and $\bar{N}$.
The proposed potential model is also consistent with the results of
partial-wave analysis of $p\bar{p}$ scattering data.

\end{document}